\documentstyle[prc,aps,preprint]{revtex}

\begin{document}
\draft
\tighten

\title{Microscopic analysis of Clausius-Duhem processes}
\author{C. Jarzynski}
\address{Theoretical Division, T-6 and T-13, MS B288 \\
         Los Alamos National Laboratory \\
         Los Alamos, NM 87545 \\
         {\tt chrisj@lanl.gov}}
\date{\today}

\maketitle

\begin{abstract}
Given a thermodynamic process which carries a system from 
one equilibrium state to another,
we construct a quantity whose average, over
an ensemble of microscopic realizations
of the process, depends only on these end states,
even if at intermediate times the system is
{\it out} of equilibrium.
This result: 
(1) can be used to express the entropy difference between
two equilibrium states in terms of an {\it irreversible}
process connecting them,
(2) leads to two statistical statements of the Clausius-Duhem 
inequality, and 
(3) can be generalized to situations in which the system 
begins and/or ends in nonequilibrium states.
\end{abstract}

\pacs{
      LAUR-98-563}

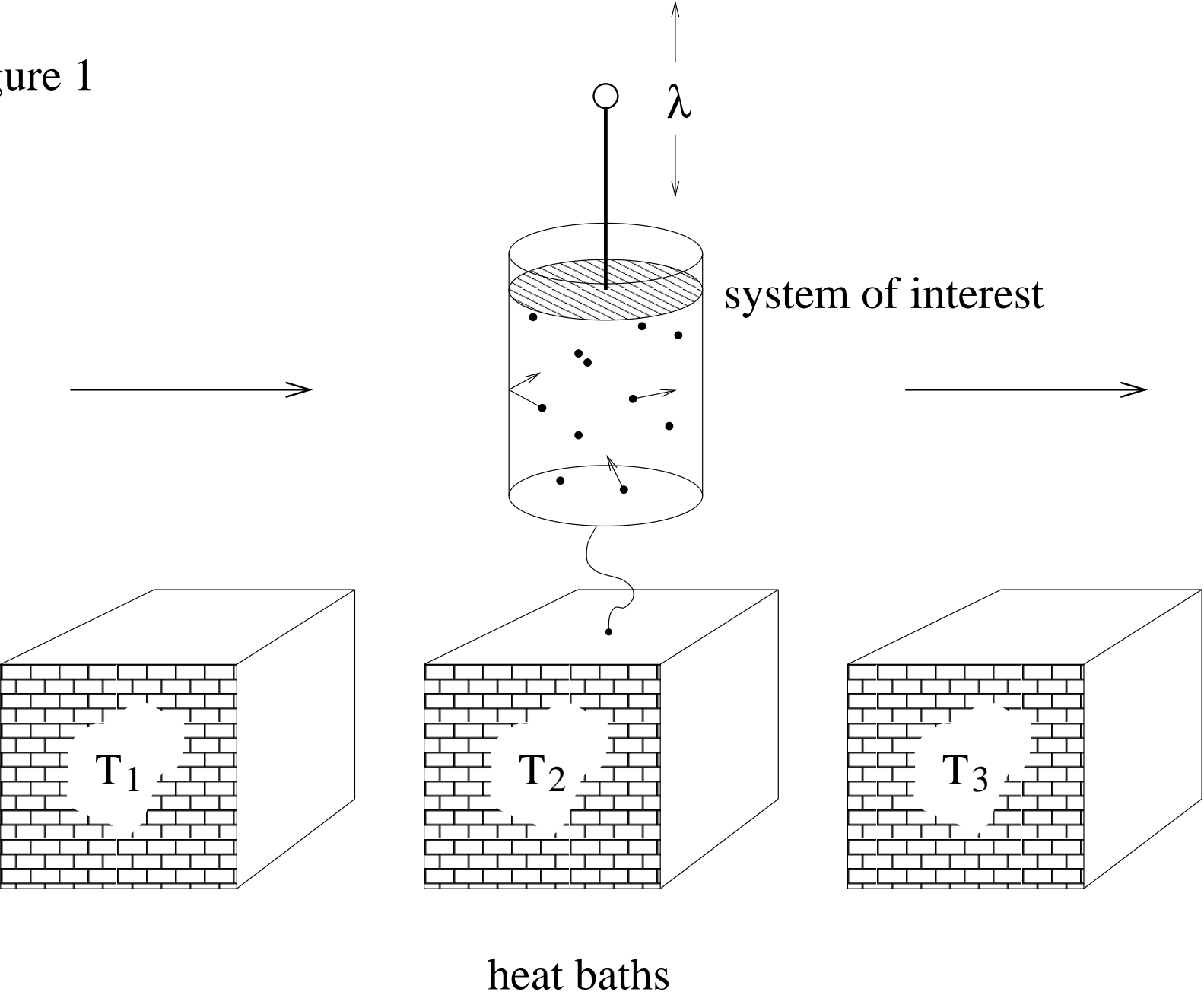
\begin{figure}
\caption{
A schematic representation of the sort of thermodynamic
process considered in this paper.
The system of interest here is a gas inside a container,
closed off at one end by a movable piston.
The position of the piston is our externally controlled
parameter, $\lambda$.
The three ``heat baths'' are simply objects with heat 
capacities much greater than that of the gas.
The system of interest is brought into thermal contact
with these baths, one at a time, e.g.\ as depicted by
the filament connecting the bath at temperature $T_2$
to the container of gas.
At the same time, $\lambda$ is varied externally.}
\label{fig:tp}
\end{figure}

The Clausius-Duhem inequality of classical thermodynamics
-- a statement of the Second Law --
applies to thermodynamic processes during which a system
evolves from one equilibrium state ($A$) to another ($B$).
It asserts that the integrated heat absorbed by the system,
divided by the temperature at which that heat
is absorbed, is bounded from above by the net change in
the entropy of the system:
\begin{equation}
\label{eq:cd}
\int_A^B {dQ\over T} \le \Delta S \equiv
S^B - S^A.
\end{equation}
By ``thermodynamic process'', we have in mind a situation
in which the system is brought into thermal contact with
a sequence of heat reservoirs at different temperatures,
one at a time, while one or more external parameters of
the system are varied with time (see Fig.\ref{fig:tp});
the denominator in Eq.\ref{eq:cd} denotes the
temperature of the reservoir from which the system
absorbs a quantity of heat $dQ$.
In general, the process is irreversible: it is carried out 
over a finite time with a finite number of reservoirs, and
the system evolves, from $A$ to $B$, through a 
sequence of {\it non-equilibrium} intermediate states.

The aim of the present paper is a classical, microscopic
analysis of such ``Clausius-Duhem'' processes, explicitly 
accounting for all degrees of freedom involved.
This analysis will follow a statistical
approach: we will consider an {\it ensemble} of
microscopic realizations of the thermodynamic process.
Each realization (described by a trajectory
specifying the evolution of all the degrees of freedom 
which make up the system of interest and reservoirs)
represents a possible ``microscopic history'', 
consistent with the macroscopic prescription for carrying 
out the thermodynamic process. 

A statistical ensemble of realizations implies
{\it fluctuations} -- from one realization to
another -- of various quantities of physical interest.
In taking a microscopic approach, therefore, 
we will be interested in the statistical {\it distribution}
(over the ensemble of microscopic histories) of values of
quantities such as $\int_A^B dQ/T$.
Working with the assumption that the system of
interest begins and ends in equilibrium states
($A\rightarrow B$), we will construct a quantity whose
average, over the ensemble of realizations,
{\it depends only on those two states}, and not on
the intermediate evolution of the system.
This result, Eq.\ref{eq:equal} below -- the central
result of this paper -- is valid even
if the system is driven far from equilibrium at
intermediate times.  
We will argue that this result can be viewed as the 
generalization -- to irreversible processes --
of the well-known identity which relates $\Delta S$ 
to an arbitrary {\it reversible} process from $A$ to $B$ 
(Eq.\ref{eq:rev}).
As we will show, the central result of this 
paper allows one to express $\Delta S$ in terms of an
arbitrary {\it irreversible} process from $A$ to $B$
(Eq.\ref{eq:deltasigirr}).
We will also show that Eq.\ref{eq:equal} immediately
implies two inequalities which are closely related to
the Clausius-Duhem inequality.
In particular, Eq.\ref{eq:probvio} places a tight upper
bound on the frequency with which finite-size violations
of the Clausius-Duhem inequality {\it do} occasionally occur.
We will finally generalize these results to
processes in which the system of
interest begins and/or ends in nonequilibrium
statistical states.

In the spirit (and level of rigor) of Gibbs\cite{gibbs}, 
we will work with the following quite general picture.
The ``system of interest'' is a closed, finite system
with a single external parameter, $\lambda$.
Additionally, we assume the presence of $N$ other
closed, finite systems,
which will play the role of heat reservoirs.
We will refer to these as ``baths'', and assume that
they have been prepared at temperatures
$T_1,\,T_2,\cdots,T_N$.
Our thermodynamic process then consists of a sequence
of steps during which the system of interest is
placed in thermal contact with the baths, one
at a time, while the value of $\lambda$
is varied along a pre-determined path $\lambda(t)$, 
over a time interval $0\le t\le\tau$.
Let $\lambda^A\equiv\lambda(0)$ and
$\lambda^B\equiv\lambda(\tau)$ denote the initial
and final parameter values, and let $n(t)$ identify
the bath with which the system is in contact at
time $t$. 
The functions $\lambda(t)$ and $n(t)$
embody the macroscopic instructions (the ``protocol'')
which specify the thermodynamic process. 

For any parameter value $\lambda$ and temperature
$T$, there exists an {\it equilibrium state}
$(\lambda,T)$ of the system of interest, described
at the microscopic level by a canonical distribution
in the phase space of the system.
Through most of the paper (up to Eq.\ref{eq:fed}),
we will restrict our attention to processes for
which the system begins ($t=0$) and ends ($t=\tau$)
in such equilibrium states, denoted by
$A\equiv(\lambda^A,T^A)$ and
$B\equiv(\lambda^B,T^B)$, respectively.
In other words, we assume that the system of interest is
prepared in equilibrium: over the
ensemble of realizations, the microscopic initial
conditions of the system are distributed canonically;
and that, at the end of the process,
the system is once again described
(statistically) by an equilibrium distribution:
the microscopic final conditions of the system
of interest are distributed canonically.
Strictly speaking, these assumptions are 
mathematically suspect:
there exists (to my knowledge) no rigorous proof that
a canonical ensemble can be achieved in a finite
time, with finite resources.
Thus, one cannot say with certainty
that there exist physically realizable methods for
preparing a system in a canonical ensemble;
or that, {\it given} a system initially in equilibrium,
there exist processes which carry that system to a
different equilibrium state in a finite time.
However, it is a widely held prejudice that such 
processes do exist in Nature, and that the canonical
ensemble is the appropriate statistical representation
for a closed system in thermal equilibrium.
We will therefore adopt the point of view that it is
a legitimate exercise to {\it assume} initial and final
equilibrium, and to explore the consequences of these
(seemingly reasonable) assumptions, 
without concerning
ourselves here with the separate problem
of establishing the validity of 
these assumptions from first principles.
Later, as mentioned, we will generalize to nonequilibrium
initial and final states.

We assume Hamiltonian evolution at the microscopic
level.
Let ${\bf z}$ denote a point in the
phase space of the system of interest --
specifying, e.g., the positions
and momenta of all its constituent particles --
and let ${\bf z}_n$ denote a point in the
phase space of the $n$th bath.
Let ${\bf y}=({\bf z},{\bf z}_1,\cdots,{\bf z}_N)$
then specify the instantaneous state of all
degrees of freedom involved.
Evolution in the full phase space (${\bf y}$-space)
is governed by a time-dependent Hamiltonian 
\begin{equation}
{\cal H}({\bf y},t) = 
H_{\lambda(t)}({\bf z}) + 
\sum_{n=1}^N H_n^b({\bf z}_n) + 
\sum_{n=1}^N \delta_{n,n(t)}
h_n^{\rm int}({\bf z},{\bf z}_n).
\end{equation}
Here, $H_\lambda({\bf z})$
is a Hamiltonian describing the system of interest,
for a parameter value $\lambda$;
$H_n^b$ is the Hamiltonian for the
$n$th bath; and
$h_n^{\rm int}$ couples the system of 
interest to that bath\cite{weakly}.
${\cal H}({\bf y},t)$ is (essentially) the most 
general classical Hamiltonian describing a system 
of interest, with a specified external-parametric time
dependence $\lambda(t)$, coupled to $N$ other
systems, one at a time.
Once the initial conditions are fully specified, 
Hamilton's equations uniquely determine 
the evolution of all degrees of freedom.

Let us consider a single realization of the
thermodynamic process, described by a 
trajectory ${\bf y}(t)$ evolving under Hamilton's
equations from some initial conditions
${\bf y}^0\equiv{\bf y}(0)$, and ending at
${\bf y}^\tau\equiv{\bf y}(\tau)$.
Let $E^0$ and $E^\tau$ denote, respectively, 
the initial and final internal energies
of the system of interest, and let
$\Delta E_n$ denote the net change in
the internal energy of the $n$th bath:
\begin{equation}
E^0 = H_{\lambda^A}({\bf z}^0) \qquad,\qquad
E^\tau = H_{\lambda^B}({\bf z}^\tau) \qquad,\qquad
\Delta E_n = H_n^b({\bf z}_n^\tau) - H_n^b({\bf z}_n^0).
\end{equation}
Now define
\begin{equation}
\label{eq:sigdef}
\Sigma \equiv {E^\tau\over T^B} - {E^0\over T^A} + 
\sum_{n=1}^N {\Delta E_n\over T_n},
\end{equation}
where $T^A$ and $T^B$ are the temperatures 
corresponding to the initial and final statistical
states of the system of interest, and the
$T_n$'s are the temperatures at which the baths
are prepared.
Working with units in which Boltzmann's constant
$k_B=1$, let us now compute 
$\langle\exp(-\Sigma)\rangle$, where {\it angular
brackets signify an average over the statistical
ensemble of realizations}, 
i.e.\ over the ensemble of trajectories ${\bf y}(t)$.
For a given realization, $\Sigma$ 
happens to depend only on the
initial and final points of ${\bf y}(t)$:
$\Sigma=\Sigma({\bf y}^0,{\bf y}^\tau)$.
Since evolution in {\bf y}-space is deterministic,
we can formally express the final conditions as 
a function of the initial conditions, 
${\bf y}^\tau = {\bf y}^\tau({\bf y}^0)$,
and then compute the desired average over 
realizations, by integrating over the distribution
of initial conditions, $f({\bf y}^0)$:
\begin{equation}
\langle\exp(-\Sigma)\rangle = 
\int d{\bf y}^0\,f({\bf y}^0)
\exp\Bigl[-\Sigma\Bigl({\bf y}^0,{\bf y}^\tau({\bf y}^0)\Bigr)\Bigr].
\end{equation}
By our assumptions regarding initial conditions,
$f({\bf y}^0)$ is a product of canonical distributions, 
hence we have
\begin{eqnarray}
\Bigl\langle \exp(-\Sigma)\Bigr\rangle &=&
{1\over{\cal N}Z^A} \int d{\bf y}^0\,
\exp \Biggl[
-{E^0\over T^A} - \sum_{n=1}^N
{H_n^b({\bf z}_n^0)\over T_n} \Biggr]
\exp(-\Sigma) \nonumber\\
\label{eq:prequal}
&=&
{1\over{\cal N}Z^A} \int d{\bf y}^\tau\,
\exp \Biggl[
-{E^\tau\over T^B} - \sum_{n=1}^N
{H_n^b({\bf z}_n^\tau)\over T_n} \Biggr] \nonumber\\
&=& {Z^B\over Z^A} =
\exp\Biggl(-{F^B\over T^B}+{F^A\over T^A}\Biggr).
\end{eqnarray}
Here, $Z^A$ and $Z^B$ denote partition
functions associated with equilibrium states
of the system of interest, and $F^A$ and $F^B$ denote
free energies:
\begin{equation}
F^i = -T^i\ln Z^i = -T^i\ln 
\int d{\bf z}\,\exp[-H_{\lambda^i}({\bf z})/T^i]
\qquad,\qquad i=A,B.
\end{equation}
${\cal N}$ is the product of the partition
functions for the $N$ baths, each corresponding
to the temperature at which that bath was prepared:
${\cal N}=\prod_{n=1}^N \int d{\bf z}_n\,
\exp[-H_n^b({\bf z}_n)/T_n]$.
In going from the first to the second line
in Eq.\ref{eq:prequal}, we have:
(1) used Eq.\ref{eq:sigdef} to 
rewrite the integrand as an explicit function
of ${\bf y}^\tau$ alone, and
(2) changed the variables of 
integration from ${\bf y}^0$ to ${\bf y}^\tau$.
By Liouville's theorem, the Jacobian 
$\vert \partial{\bf y}^\tau/\partial{\bf y}^0\vert = 1$.

At this point, Eq.\ref{eq:prequal} is just a statement
about energy exchange among a number of finite, closed
systems, one of which has been singled out as being 
``of interest''.
Its validity does not depend on the relative sizes
of these objects.
To establish contact with more familiar results,
let us now imagine the limiting case in which the
heat capacities of the $N$ baths become arbitrarily greater 
than that of the system of interest.
The baths then assume the role of ``infinite'' heat
reservoirs, and we can rewrite the term 
$\sum_n \Delta E_n/T_n$
in Eq.\ref{eq:sigdef} as $-\int_A^B dQ/T$:
\begin{equation}
\Sigma = {E^\tau\over T^B} - {E^0\over T^A} -
\int_A^B {dQ\over T}.
\end{equation}
(During the time interval over which the 
system of interest is coupled to the $n$'th bath, the 
heat absorbed by the system is equal to the energy
lost by that bath, so the value of $\int dQ/T$ over
that interval of time is simply $-\Delta E_n/T_n$.)
Eq.\ref{eq:prequal} now becomes:
\begin{equation}
\label{eq:equal}
\Biggl\langle\exp\Biggl[
-\Delta\Biggl({E\over T}\Biggr) + 
\int_A^B {dQ\over T}\Biggr]\Biggr\rangle = 
\exp \Biggl[-\Delta\Biggl({F\over T}\Biggr)\Biggr],
\end{equation}
using the shorthand notation 
$\Delta(E/T)\equiv E^\tau/T^B-E^0/T^A$ and
$\Delta(F/T)\equiv F^B/T^B-F^A/T^A$.
This is the central result of this paper.
We add here that a result equivalent to
Eq.\ref{eq:equal} has been derived independently
by Gavin E.\ Crooks\cite{crooks} -- using
stochastic, Markovian dynamics to model the
evolution of the system of interest --
and that this result has been shown
to be closely related to the Fluctuation Theorem
for non-equilibrium steady states.\cite{ecm,es,gc}

Since internal energies are quantities
associated with specific microscopic states
(i.e.\ points in phase space), the
exact values of $E^0$, $E^\tau$ and
$\int_A^B dQ/T\,(=-\sum_n\Delta E_n/T_n$)
differ from one realization of the thermodynamic
process to the next. 
By contrast, the free energies $F^A$ and
$F^B$ are associated with {\it canonical ensembles}
of microstates of the system of interest. 
Thus, $\Sigma=\Delta(E/T)-\int_A^B dQ/T$ 
is a linear combination of quantities 
($E^0$, $E^\tau$, the $\Delta E_n$'s) 
which vary in value from one 
realization to the next, and Eq.\ref{eq:equal}
makes an assertion regarding the statistical
distribution of values of $\Sigma$:
it claims that the average of $\exp(-\Sigma)$,
over the ensemble of realizations, 
is equal to $\exp[-\Delta(F/T)]$, 
which depends only on the equilibrium states
$A$ and $B$, and not on the sequence of (nonequilibrium)
statistical states through which the system evolves
in getting from $A$ to $B$!
Now, for macroscopic, {\it reversible} processes, 
it is well known that\footnote{
Since quantities such as heat, entropy, etc.,
appear in both thermodynamics and statistical
mechanics, and since these are (formally)
separate theories, it is useful to distinguish
between the two.
Here and below, we use a carat (e.g. $d\hat Q$) to 
denote that a certain quantity is to be understood 
in the macroscopic (thermodynamic) context, rather 
than in the microscopic (statistical) context.}
\begin{equation}
\label{eq:rev}
\int_A^B {d\hat Q\over T} = \Delta \hat S 
\equiv \hat S^B - \hat S^A
\qquad ({\rm REVERSIBLE}),
\end{equation}
regardless of the path (through equilibrium state
space) taken from $A$ to $B$.
Since Eq.\ref{eq:rev} can be written, at the 
macroscopic level, as
$\hat\Sigma = \Delta(\hat F/T)$
(because $\hat F=\hat E-\hat ST$), our central result
[$\langle e^{-\Sigma}\rangle
=e^{-\Delta(F/T)}$]
may be viewed as the microscopic extension of 
Eq.\ref{eq:rev} to {\it irreversible} processes.
Moreover, and somewhat surprisingly, this result
is valid {\it regardless of how far the system is
driven away from equilibrium} between the initial
and final times:
no matter how violent the process, Eq.\ref{eq:equal}
will hold, provided the system begins in $A$ and 
ends in $B$.

For a given equilibrium state, the microscopic
expressions for free energy, average internal energy, 
and entropy satisfy $F=\overline{E}-ST$, where the
overbar denotes an equilibrium (canonical) average.
Thus, the quantity which we have called 
$\Delta(F/T)$ can be written as:
\begin{equation}
\Delta\Biggl({F\over T}\Biggr) = 
{\overline{E}^B\over T^B} - {\overline{E}^A\over T^A}
-\Delta S =
{\langle E^\tau\rangle\over T^B} -
{\langle E^0\rangle\over T^A} - \Delta S,
\end{equation}
where the second equality follows from our assumptions
regarding the initial and final distributions of
microstates.
Combining this with Eq.\ref{eq:equal} gives:
\begin{equation}
\label{eq:deltasigirr}
\Delta S = 
{\langle E^\tau\rangle\over T^B} -
{\langle E^0\rangle\over T^A}
+ \ln 
\Biggl\langle \exp 
\Biggl[ -\Delta \Biggl({E\over T}\Biggr) + 
\int_A^B{dQ\over T} \Biggr]\Biggr\rangle.
\end{equation}
This result expresses the entropy difference
$\Delta S=S^B-S^A$ in terms of an arbitrary --
in general {\it irreversible} -- thermodynamic process
from $A$ to $B$.
In principle, by repeatedly measuring $E^0$, $E^\tau$,
and $\int_A^B dQ/T$ for independent realizations of such
a process, we can construct the averages appearing in
Eq.\ref{eq:deltasigirr}, and therefore compute the
value of $\Delta S$.\footnote{
Note that Eq.\ref{eq:deltasigirr} generally involves 
averaging over infinitely many finite-time realizations, 
in contrast to Eq.\ref{eq:rev}, which gives $\Delta S$ in
terms of a single realization of infinite duration.}

We now derive, as a byproduct of Eq.\ref{eq:equal},
two inequalities, one old and one new (Eqs.\ref{eq:statcd}
and \ref{eq:probvio} below), which are closely related
to the Clausius-Duhem inequality. 
By the convexity of the function
$e^x$, Eq.\ref{eq:equal} implies
\begin{equation}
\label{eq:ineq}
-{\langle E^\tau\rangle\over T^B} 
+{\langle E^0\rangle\over T^A} 
+ \Biggl\langle\int_A^B {dQ\over T}\Biggr\rangle
\le - {F^B\over T^B} + {F^A\over T^A}.
\end{equation}
Once again invoking the identity $F=\overline{E}-ST$,
along with our assumptions of initial and final
equilibria, $\langle E^0\rangle=\overline{E}^A$ and
$\langle E^\tau\rangle=\overline{E}^B$,
we can rewrite Eq.\ref{eq:ineq} as:
\begin{equation}
\label{eq:statcd}
\Biggl\langle \int_A^B {dQ\over T} \Biggr\rangle
\le \Delta S.
\end{equation}
This result says, effectively, that the Clausius-Duhem 
inequality
is satisfied ``on average'', where the average is taken 
over an ensemble of microscopic realizations of a 
given thermodynamic process.
This still leaves open the possibility that
there exist individual realizations for which the 
inequality is violated.
We will now use Eq.\ref{eq:equal} to investigate 
the frequency of occurrence of such violations.

Macroscopically, the Clausius-Duhem 
inequality can be written as
$\Delta(\hat E/T)-\int_A^B d\hat Q/T\ge\Delta(\hat F/T)$, 
or simply $\hat\Sigma\ge\Delta(\hat F/T)$.
Thus, thermodynamics tells us that we
will ``never'' observe a value of
$\hat\Sigma$ below $\Delta(\hat F/T)$.
To investigate the {\it microscopic} validity of this 
statement, let $p(\Sigma)$ denote the distribution of 
values of $\Sigma$ corresponding to the statistical
ensemble of microscopic realizations of a given
thermodynamic process.
Then the probability of observing a value of
$\Sigma$ no greater than some fixed 
value $\Sigma_0$ is just:
${\rm Prob}[\Sigma\le\Sigma_0]
= \int_{-\infty}^{\Sigma_0}d\Sigma\,p(\Sigma)$.
But Eq.\ref{eq:equal} tells us that
$\int_{-\infty}^{+\infty}p e^{-\Sigma}=
e^{-\Delta(F/T)}$.
When we combine this with the inequality
chain
\begin{equation}
\int_{-\infty}^{+\infty} p e^{-\Sigma}
\ge \int_{-\infty}^{\Sigma_0} p e^{-\Sigma}
\ge e^{-\Sigma_0}\int_{-\infty}^{\Sigma_0} p,
\end{equation}
and take $\Sigma_0=\Delta(F/T)-\Gamma_0$, 
where $\Gamma_0>0$, we obtain
\begin{equation}
\label{eq:probvio}
{\rm Prob}[\Sigma\le\Delta(F/T)-\Gamma_0]
\le \exp(-\Gamma_0/k_B),
\end{equation}
where we have explicitly put in the Boltzmann
constant $k_B$.
Thus, {\it the probability
of observing a violation of the Clausius-Duhem
inequality, by an amount no less than 
$\Gamma_0$, is bounded from above by}
$e^{-\Gamma_0/k_B}$.
A macroscopic violation would be one for which
$\Gamma_0/k_B\gg 1$, hence 
such violations are extremely rare: 
the Clausius-Duhem inequality is ``never'' violated by a
macroscopic amount\footnote{
It is interesting to note the similarity between
Eq.\ref{eq:probvio} -- which pertains to a nonequilibrium
thermodynamic process -- and the Einstein-Boltzmann expression 
for microscopic fluctuations of a system in equilibrium; 
see e.g.\ Ref.\cite{ll}, Eq.[112.1].}.

The previous two paragraphs are by no means intended
as a first-principles derivation of the Second Law,
as they assume -- reasonably, but without proof -- 
canonical distributions.
Indeed, it has long been known
(see, e.g.\ Ref.\cite{gibbs}) that canonical 
ensembles imply inequalities such as Eq.\ref{eq:statcd}, 
stating that the Second Law is not violated ``on average''.
(However, I believe that the upper bound given by
Eq.\ref{eq:probvio} is a new result.)
The aim here is rather to reveal the close connection
between the central result of this paper (Eq.\ref{eq:equal}) 
and the Clausius-Duhem inequality. 

While Eq.\ref{eq:equal} is valid for any thermodynamic 
process which carries a system from $A$ to $B$, it is 
instructive to ponder limiting cases of such processes.
We will now consider three examples, for which we will
be able to verify Eq.\ref{eq:equal} directly (without 
invoking Liouville's theorem), by solving explicitly for 
$\Sigma$.

The first example involves bringing a system -- initially 
at a temperature $T^A$ -- into contact with a reservoir 
at temperature $T^B$, and allowing the system to relax to
the temperature of the reservoir. 
Let us therefore imagine that, at time $t=0$, we start 
with the system in the equilibrium state $A=(\lambda,T^A)$.
Then, at $t=0^+$ (``immediately after $t=0$''),
we place the system in thermal contact with a reservoir
at temperature $T^B$, and let the two equilibrate.
We assume the reservoir to have the usual property of
``infinite'' heat capacity, so that the system of
interest relaxes to the equilibrium state
$B=(\lambda,T^B)$.
In this situation, we get
$\int_A^B dQ/T = (1/T^B) \int_A^B dQ = (E^\tau-E^0)/T^B$ --
where the initial and final energies are given by
$E^0=H_\lambda({\bf z}^0)$ and 
$E^\tau=H_\lambda({\bf z}^\tau)$ --
from which it follows that
\begin{equation}
\Sigma = {E^\tau\over T^B} - {E^0\over T^A} -
{E^\tau-E^0\over T^B} = 
{H_\lambda({\bf z}^0) \over T^B} -
{H_\lambda({\bf z}^0) \over T^A}.
\end{equation}
We now average over realizations by integrating over
the distribution of initial conditions to get:
\begin{eqnarray}
\Bigl\langle \exp(-\Sigma)\Bigr\rangle &=&
{1\over Z^A} \int d{\bf z}^0
\exp\Biggl[-{H_\lambda({\bf z}^0)\over T^A}\Biggr]
\exp(-\Sigma) \\
&=& {1\over Z^A} \int d{\bf z}^0 
\exp\Biggl[-{H_\lambda({\bf z}^0)\over T^B}\Biggr] \\ 
&=& {Z^B\over Z^A} = \exp\Biggl[-\Delta
\Biggl({F\over T}\Biggr)\Biggr],
\end{eqnarray}
in agreement with Eq.\ref{eq:equal}.

The second example involves making a sudden change in the
value of the external parameter, $\lambda^A\rightarrow\lambda^B$,
and then letting the system (assumed in contact at all times
with a reservoir at temperature $T$) relax to the equilibrium
state corresponding to the new parameter value.
Thus, we begin ($t=0$) with the system in 
equilibrium state $A=(\lambda^A,T)$, coupled to a
reservoir at temperature $T$;
a moment later ($t=0^+$), we instantaneously 
change the parameter value from $\lambda^A$ to 
$\lambda^B$, and then we allow the system to relax to the
equilibrium state $B=(\lambda^B,T)$.
The initial energy of the system 
is given by $E^0=H_{\lambda^A}({\bf z}^0)$;
the energy just after $\lambda$ is switched
to $\lambda^B$ is given by
$E^{0^+}=H_{\lambda^B}({\bf z}^0)$; and the
final energy is $E^{\tau}=H_{\lambda^B}({\bf z}^\tau)$.
Then $\int_A^B dQ/T = (E^\tau-E^{0^+})/T$, and
\begin{equation}
\Sigma = {E^\tau\over T} - {E^0\over T} -
{E^\tau - E^{0^+} \over T}
= {H_{\lambda^B}({\bf z}^0) \over T} - 
{H_{\lambda^A}({\bf z}^0) \over T},
\end{equation}
from which we again get
\begin{equation}
\Bigl\langle \exp(-\Sigma)\Bigr\rangle =
{1\over Z^A} \int d{\bf z}^0
\exp\Biggl[-{H_{\lambda^A}({\bf z}^0)\over T}\Biggr]
\exp(-\Sigma) = 
\exp\Biggl[-\Delta\Biggl({F\over T}\Biggr)\Biggr].
\end{equation}

The third example combines these two, and gives us
a specific prescription for carrying a system from
one arbitrary equilibrium state to another.
We start with the system in state
$A=(\lambda^A,T^A)$.
Then we instantaneously switch the parameter value
to $\lambda^B$, after which we
place the system in contact with a reservoir at
temperature $T^B$, and allow it to relax to the
state $B=(\lambda^B,T^B)$.
Following steps as above, we get
$\Sigma = H_{\lambda^B}({\bf z}^0)/T^B -
H_{\lambda^A}({\bf z}^0)/T^A$, from which it 
once more immediately follows that 
$\langle e^{-\Sigma}\rangle = e^{-\Delta(F/T)}$.

In each of these examples, a convenient 
cancellation of terms gave us the value of $\Sigma$
explicitly in terms of known functions of the inital 
conditions of the system of interest (${\bf z}^0$).
For a more general thermodynamic process, in
which more reservoirs are involved and $\lambda$
changes at a finite rate, this is not the case;
if we wanted to compute $\Sigma$ for a realization
launched from a known set of initial conditions,
we would need to actually integrate the equations
of motion (in the {\it full} phase space) to get
$E^\tau$ and $\int_A^B dQ/T$.

Let us now suppose that we prepare the system in state
$A=(\lambda^A,T)$, and we switch the parameter at 
an arbitrary {\it finite} rate from $\lambda^A$ to 
$\lambda^B$ (driving the system out of equilibrium),
while keeping the system in thermal contact with a reservoir
at temperature $T$;
at the end we hold $\lambda$ fixed at $\lambda^B$
and allow the system to relax to $B=(\lambda^B,T)$.
In this situation, we have
\begin{equation}
\int_A^B {dQ\over T} = 
{1\over T}\int_A^B dQ = 
{1\over T} (E^\tau-E^0-W),
\end{equation}
where $W$ is the external work performed on the system
by driving the parameter.
We thus have $\Sigma=W/T$, and Eq.\ref{eq:equal} reduces
to the following relationship between the work
performed (during realizations of this nonequilibrium
process) and the free energy difference
$\Delta F \equiv F^B-F^A$:
\begin{equation}
\label{eq:fed}
\Biggl\langle\exp\Biggl(-{W\over T}\Biggr)\Biggr\rangle
= \exp\Biggl(-{\Delta F\over T}\Biggr).
\end{equation}
Note the conditions for the validity of this result:
there is only one heat reservoir\footnote{
Or, if there are more, they share a common temperature.}
and its temperature must equal that at which the system 
is initially prepared.
Eq.\ref{eq:fed} has recently been derived in a number
of ways, and confirmed in numerical
experiments\cite{fe}.

We now generalize our analysis to the situation in which 
the system of interest begins and ends in nonequilibrium
statistical states.
For instance, we might prepare the system by heating
it at one end and cooling at another, until a steady-state
thermal gradient is acheived.
(The $N$ baths, however, are prepared in equilibrium, 
as before.)
Whatever the method of preparation, let $\rho^0({\bf z})$ 
denote the statistical distribution of initial conditions 
(of the system of interest) achieved by that preparation.
Similarly, let $\rho^\tau({\bf z})$ denote 
the distribution of final conditions; this will 
of course depend on the sequence of steps defining the 
thermodynamic process. 

As before, a realization of the process 
is described by a microscopic trajectory ${\bf y}(t)$,
determined by the initial conditions, ${\bf y}^0$.
For a given realization, let us define
\begin{equation}
\Gamma \equiv -\ln\rho^\tau({\bf z}^\tau)
+\ln\rho^0({\bf z}^0) + 
\sum_{n=1}^N {\Delta E_n\over T_n},
\end{equation}
and let us compute the average of 
$\exp(-\Gamma)$ over our ensemble of
realizations:
\begin{equation}
\label{eq:nequal}
\langle\exp(-\Gamma)\rangle =
{1\over{\cal N}} \int d{\bf y}^0\,
\rho^0({\bf z}^0) 
\exp \Biggl[ - \sum_{n=1}^N
{H_n^b({\bf z}_n^0)\over T_n} \Biggr]
\exp(-\Gamma) = 1,
\end{equation}
following steps like those leading to
Eq.\ref{eq:prequal}.

Since we were able to derive 
Eq.\ref{eq:statcd} from Eq.\ref{eq:equal},
it is natural to wonder whether an interesting
inequality can similarly be obtained from
Eq.\ref{eq:nequal}.
The convexity of $e^x$ in this case gives us
$-\langle\Gamma\rangle\le 0$, or
\begin{equation}
\label{eq:prineqs}
\langle\ln\rho^\tau({\bf z}^\tau)\rangle
-\langle\ln\rho^0({\bf z}^0)\rangle +
\langle\smallint dQ/T\rangle
\le 0.
\end{equation}
Now note that 
$-\langle\ln\rho^0({\bf z}^0)\rangle =
-\int\rho^0\ln\rho^0 = S_G[\rho^0]$,
where the integration
is over the phase space of the system of interest,
and $S_G[\rho^0]$ represents the statistical (Gibbs)
entropy associated with the
initial statistical state of the system.
Similarly, $-\langle\ln\rho^\tau({\bf z}^\tau)\rangle
=S_G[\rho^\tau]$.
Eq.\ref{eq:prineqs} then reads
\begin{equation}
\label{eq:nneq}
\langle\smallint dQ/T\rangle
\le S_G[\rho^\tau] - S_G[\rho^0]
\equiv\Delta S_G.
\end{equation}
That is, {\it the expectation value of 
$\int dQ/T$
is bounded from above by the net change
in the statistical entropy}, $S_G$, characterizing
the initial and final states of the system of interest.
Note that in the case of an isolated system (no heat
baths), this inequality reduces to a trivial result,
as both sides are identically zero.

Eq.\ref{eq:nneq} is equivalent to the
statement: $\langle\Gamma\rangle\ge 0$.
Following a line of reasoning like the
one leading to Eq.\ref{eq:probvio}, 
we can use Eq.\ref{eq:nequal} to place an
upper bound on the probability of observing 
a value of $\Gamma$ no greater than $-\Gamma_0$:
\begin{equation}
\label{eq:nprobvio}
{\rm Prob}[\Gamma\le-\Gamma_0]\le e^{-\Gamma_0/k_B}.
\end{equation}
Thus, we will ``never'' observe a 
``macroscopically negative'' value of $\Gamma$.
Eqs.\ref{eq:nequal}, \ref{eq:nneq} and \ref{eq:nprobvio} 
together constitute a generalization of 
Eqs.\ref{eq:equal}, \ref{eq:statcd} and \ref{eq:probvio},
to situations in which the system of interest
begins and ends
in states not necessarily corresponding to
thermal equilibrium.

A macroscopic, reversible process 
between two equilibrium states $A$ and $B$ has the
property that 
$\Delta(\hat E/T)-\int_A^B d\hat Q/T=\Delta(\hat F/T)$ 
(Eq.\ref{eq:rev}).
The central result of this paper is a 
microscopic, statistical generalization
of this result to {\it irreversible} 
processes between two such states:
$\langle e^{-\Delta(E/T)+\int_A^B dQ/T}\rangle=
e^{-\Delta(F/T)}$ (Eq.\ref{eq:equal}),
where the average is taken over an ensemble of
realizations of the process.
In both cases, the right side of the equation
depends only on the states $A$ and $B$, 
and not on the (equilibrium or nonequilibrium) path
connecting them.
We have used Eq.\ref{eq:equal} to derive an 
expression for the entropy difference between two
equilibrium states, in terms of an arbitrary
(generally irreversible) thermodynamic process
connecting them (Eq.\ref{eq:deltasigirr}).
We have also shown that Eq.\ref{eq:equal} leads to
statistical statements of the Clausius-Duhem inequality,
in particular placing an upper bound on 
the probability for observing violations
of the Clausius-Duhem inequality above an arbitrary
threshold $\Gamma_0$ (Eq.\ref{eq:probvio}).
Finally, we have extended this analysis to
processes which begin and/or end out of equilibrium.

\section*{Acknowledgments}

I would like to thank G.Crooks, E.Lieb, C.Maes, and 
J.Percus for stimulating conversations and correspondence 
regarding central issues in this paper, and C.Zalka for 
simplifying the derivation of Eq.\ref{eq:probvio}.
This research was partially supported by the
Polish-American Maria Sk\l odowska-Curie Joint
Fund II, under project PAA/NSF-96-253.

\begin{references}
\bibitem{weakly}
As usual, we take the interaction energies $h_n^{\rm int}$
to be negligible in comparison with the other terms
in ${\cal H}$. 
While this assumption does not enter explicit 
calculations, it allows us to view the total energy
of the system of interest and $N$ baths as simply
the sum of the internal energies of these $N+1$ objects.
The terms appearing in the First Law, 
$dE=dW+dQ$, are then unambiguously defined.
\end{references}
\end{document}